\documentclass[aps,prl,twocolumn,showpacs]{revtex4}
\usepackage{amsfonts,amssymb,amsmath,latexsym,times} 
\usepackage[dvips]{graphics}

\begin{document}

[Phys. Rev. Lett. 101, 054102 (2008)]

\title{Calculation of Superdiffusion for the Chirikov-Taylor Model}

\author{Roberto Venegeroles}
\email{roberto.venegeroles@ufabc.edu.br}
\affiliation{Centro de Matem\'atica, Computa\c c\~ao e Cogni\c c\~ao, Universidade Federal do ABC, 09210-170, Santo Andr\'e, SP, Brazil}

\date{\today}

\begin{abstract}

It is widely known that the paradigmatic Chirikov-Taylor model presents enhanced diffusion for specific intervals of its stochasticity parameter due to islands of stability, which are elliptic orbits surrounding accelerator mode fixed points. In contrast to the normal diffusion, its effect has never been analytically calculated. Here we introduce a differential form for the Perron-Frobenius evolution operator in which normal diffusion and superdiffusion are treated separately through phases formed by angular wavenumbers. The superdiffusion coefficient is then calculated analitically resulting in a Schloemilch series with an exponent $\beta=3/2$ for the divergences. Numerical simulations support our results.
\end{abstract}

\maketitle

One of the most fundamental questions in the study of transport theory is to understand strange diffusion, whereby a mean-squared displacement follows an asymptotic time power law $\left\langle x^{2}\right\rangle\sim t^{\beta}$ for $\beta\neq1$, starting from nonlinear microscopic equations of dynamics. The well-known Brownian (normal) diffusion is characterized by the exponent $\beta=1$, while the strange diffusion is subdivided into two types: superdiffusive ($\beta>1$) and subdiffusive ($\beta<1$). The strange transport has been studied and observed in a wide range of systems such as plasmas \cite{plasma}, flows and turbulence \cite{flow}, amorphous semiconductors, porous media, glasses, granular matter \cite {disordered}, one-dimensional intermittent maps \cite{1D}, and biological cell motility \cite{bio}.

In this Letter, superdiffusion is studied analytically in the paradigmatic Chirikov-Taylor standard map:
\begin{eqnarray} 
\begin{array}{l} I_{n+1} = I_{n}+K\sin\theta_{n},\\ 
                 \theta_{n+1} = \theta_{n}+I_{n+1}\qquad \mbox{mod}\, 2\pi,
\end{array} \label{map} 
\end{eqnarray}
defined on the cylinder $-\pi\leq\theta<\pi$, $-\infty<I<\infty$, where $K$ is the stochasticity parameter. In the past, many authors investigated the diffusive behavior of chaotic systems on the basis of map (\ref{map}) through the diffusion coefficient $D$, defined as
\begin{equation}
\label{diffdef}
D=\frac{\left\langle (I_{n}-I_{0})^{2}\right\rangle}{2n}\sim n^{\beta-1},\qquad n\gg1. 
\end{equation}
Brownian diffusion emerges when the last {\it KAM} barrier is destroyed at the critical parameter value $K_{c}\approx0.9716$ \cite{Greene}. The corresponding diffusion coefficient $D(K)$ exhibits an oscillatory feature for $K>K_{c}$. This behavior was first numerically observed by Chirikov \cite{Chirikov} and subsequently calculated by Rechester {\it et al.} \cite{Rechester}. In addition to the normal diffusion, the map (\ref{map}) also exhibits superdiffusion of the L\'evy type ($1<\beta<2$) due to accelerator mode islands, which are created around a stable $Q$-periodic orbit $\left\{(I_{n}^{*},\theta_{n}^{*})\right\}_{1\leq n\leq Q}$, satisfying \cite{Ishizaki}
\begin{eqnarray}
\label{accmod}
I_{n+Q}^{*}-I_{n}^{*}=2l\pi,\qquad K\sum_{n=1}^{Q}\sin\theta_{n}^{*}=2l\pi,
\end{eqnarray}
for any integer $l\neq0$. For $Q=1$ we have the following stability condition \cite{Lichtenberg}:
\begin{equation}
\label{stab}
\mathcal{S}_{\theta}: |2+K\cos\theta|<2.
\end{equation}
Typically, trajectories diffuse normally, although some of them may be dragged along the accelerated modes. Such coexistence between chaotic and regular orbits results in a complex dynamics whose exponent is certainly less than the ballistic value $\beta=2$. This fact has motivated the use of suitable statistical theories such as continuous-time random walk (CTRW) \cite{Zumofen} and fractional kinetics (FK) \cite{Zaslavsky} in the numerical study of the superdiffusion. Here, the superdiffusion coefficient and its corresponding transport exponent are analytically calculated from first principles, i.e., from dynamics ruled by Eq. (\ref{map}).

The analysis of the map (\ref{map}) is best carried out in Fourier space. The Fourier expansion of conditional probability density $\rho_{n}$, in which an initial state ($I_{0},\theta_{0}$) at $n=0$ evolves to a final state $(I,\theta)$ at step $n$, can be written in the form
\begin{equation}
\label{distribution}
\rho_{n}(I,\theta|I_{0},\theta_{0})=\sum_{m}e^{im\theta}\tilde{a}_{n}(m,I).
\end{equation}
Relations between normal diffusion and Pollicott-Ruelle resonances for a general class of area-preserving maps that includes Eq. (\ref{map}) as a particular case was studied in Ref. \cite{Venegeroles} through the evolution of transformed amplitudes $a_{n}(m,q)=(2\pi)^{-1}\int dI\,\tilde{a}_{n}(m,I)e^{-iqI}$. On the other hand, as it will be seen latter, the ampltudes $\tilde{a}_{n}(m,I)$ are proving to be more suitable in the description of superdiffusion. The discrete time evolution of these amplitudes are also ruled by the Perron-Frobenius (PF) operator $\hat{U}$, defined as $\tilde{a}_{n}(m,I)=\hat{U}\tilde{a}_{n-1}(m,I)$. More explicitly, starting from the evolution of $a_{n}(m,q)$ we have
\begin{eqnarray}
\label{ainvert}
a_{n}(m,q)&=&(2\pi)^{-1}\int d\theta dI\,e^{-i(m\theta+qI)}\nonumber\\
&\,&\times\int d\theta'dI'\delta(I-I'-K\sin\theta')\delta(\theta-\theta'-I)\nonumber\\
&\,&\times\sum_{m'}\int dq'\,e^{i(m'\theta'+q'I')}\,a_{n-1}(m',q').
\end{eqnarray}
The useful PF operator for $\tilde{a}_{n}(m,I)$ can be obtained in three steps: $i$) integrating Eq. (\ref{ainvert}) on $\theta$ and $I'$; $ii$) multiplying the two sides of the resulting equation by $\exp(iqI')$ and integrating both on $q$; and $iii)$ considering the translation operator $\tilde{a}(I+x)=\exp(x\partial_{I})\tilde{a}(I)$. Finally, the dynamics of the system (\ref{map}) induces the following law of evolution
\begin{eqnarray}
\label{evolution}
\tilde{a}_{n}(m,I)=e^{-imI}\sum_{m'}J_{m-m'}(iK\partial_{I})\tilde{a}_{n-1}(m',I),
\end{eqnarray}
where $J_{m}(x)$ is the Bessel function of the first kind. Hereafter, we will use the abbreviation $J_{m}(iK\partial_{I})\equiv\hat{J}_{m}$. The initial probability amplitude is given by $\tilde{a}_{0}(m,I)=(2\pi)^{-1}e^{-im\theta_{0}}\delta(I-I_{0})$, which represents the certainty of an arbitrary initial condition $I=I_{0}$ and $\theta=\theta_{0}$ at $n=0$. Besides, we must define the density to generate initial values $(I_{0},\theta_{0})$, which are independent from the dynamics. Usually, the initial values are assumed as uniformly distributed in their respective domains \cite{n1}. In this sense, numerical calculations of (\ref{diffdef}) are performed as $D_{num}=N^{-1}\sum_{j=1}^{N}(I_{n,j}-I_{0,j})^{2}/2n$, where $N\gg1$ is the number of random initial conditions generated for each value of $K$.

In order to calculate the superdiffusion coefficient $D_{sup}$ it is useful to calculate the normal diffusion coeffcient $D_{nor}$ by means of this new approach. In this context, arbitrary moments depend only on the state $m=0$ at instant $n$: $\left\langle I^{p}\right\rangle=(2\pi)^{2}[(i\partial_{q})^{p}a_{n}(0,q)]_{q=0}=(2\pi)\int dI\,\tilde{a}_{n}(0,I)I^{p}$. Thus, we can decompose the projected PF operator $P\hat{U}^{n}$ into two parts: $P\hat{U}^{n}=P\hat{U}^{n}P+P\hat{U}^{n}Q$, where $P$ and $Q$ are mutually orthogonal projection operators that represent the states $m=0$ and $m\neq0$, respectively. On the other hand, $P\hat{U}^{n}Q$ can be neglected due to $\tilde{a}_{0}(m\neq0,q)\propto e^{-im\theta_{0}}$, whose expected value disappears at random initial conditions on $[-\pi,\pi)$. Hence, $P\hat{U}^{n}P$ rules the diffusion in the normal regime \cite{Venegeroles}. Now consider the leading case where $m_{1}=m_{2}=\ldots=m_{n}=0$, which gives $\hat{U}^{n}=\hat{J}_{0}^{n}=1+n(K^{2}/4)\partial_{I}^{2}+\mathcal{O}(\partial_{I}^{4})$ and the quasilinear diffusion $D_{ql}=K^{2}/4$. Corrections to this value are obtained by the introduction of a small $k$-piece of wavenumbers in a quasilinear stripe of length $n$, as shown in the example below for $k=3$:
\begin{equation}
\label{string1}
\overbrace{(0,0)\ldots\underbrace{(0,m_{1})(m_{1},m_{2})(m_{2},m_{3})(m_{3},0)}_{k=3}\ldots(0,0)}^{n}.
\end{equation}
Considering $k\ll n$ we can neglect border effects, so that we have $n$ configurations of the type (\ref{string1}) that gives the same result for $P\hat{U}^{n}P$. Adding to these the quasilinear configuration we have
\begin{equation}
\label{at}
\tilde{a}_{n}(0,I)=\frac{1}{2\pi}\left(1+n\sum_{m_{1}}\sum_{m_{2}}\sum_{m_{3}}\hat{\Omega}\right)\,\hat{J}_{0}^{n}\,\delta(I-I_{0}),
\end{equation}
where the operator $\hat{\Omega}$ is given by
\begin{eqnarray}
\label{omega}
\hat{\Omega}&=&\hat{J}_{-m_{1}}\times e^{-i(m_{1}+m_{2}+m_{3})I}\nonumber\\
&\times&[J_{m_{1}-m_{2}}(m_{2}K)+J_{m_{1}-m_{2}}(m_{3}K)+\hat{J}_{m_{1}-m_{2}}]\nonumber\\
&\times&[J_{m_{2}-m_{3}}(m_{3}K)+\hat{J}_{m_{2}-m_{3}}]\times\hat{J}_{m_{3}}.
\end{eqnarray}
Thus, in order to obtain only relevant contributions to the diffusion coefficient, we should perform summation in Eqs. (\ref{at},\ref{omega}) over all integers subject to the constraint $m_{1}+m_{2}+m_{3}=0$. Besides, Bessel operators follow the same series expansions of Bessel functions, namely $\hat{J}_{m}\propto\partial_{I}^{m}[1+\mathcal{O}(\partial_{I}^{2})]$. Evidently, wavenumbers that contribute with terms $\mathcal{O}(\partial_{I}^{3})$ can be neglected in the calculation of the moment $\left\langle I^{2}\right\rangle$ resulting, in addition to $m_{3}=-(m_{1}+m_{2})$, in the three independent $j$-constraints
$|m_{1}|+|\mu_{j}m_{1}+\sigma_{j}m_{2}|+|m_{1}+m_{2}|\leq2$, where
$(\mu_{j},\sigma_{j})=\left\{(0,0),(1,-1),(1,2)\right\}$. Applying these
constraints, the sum $\sum_{m_{1},m_{2},m_{3}}\hat{\Omega}$ becomes $\hat{\Omega}_{1}+\hat{\Omega}_{2}$ where
\begin{eqnarray}
\hat{\Omega}_{1}&=&2J_{2}(K)\hat{J}_{0}\hat{J}_{1}^{2},\\
\hat{\Omega}_{2}&=&2J_{1}^{2}(K)\hat{J}_{1}^{2}+4J_{2}(2K)J_{4}(2K)\hat{J}_{0}\hat{J}_{2}+J_{3}(K)\nonumber\\
&\times&[-J_{3}(2K)+J_{3}(K)+J_{1}(2K)-J_{1}(K)]\hat{J}_{1}^{2}.\nonumber\\
\end{eqnarray}
Finally, the normal diffusion coefficient is calculated as
\begin{eqnarray}
\label{dnor}
\frac{D_{nor}}{D_{ql}}&=&\frac{1}{2n}\int d\mathcal{I}[1+n(\hat{\Omega}_{1}+\hat{\Omega}_{2})](D_{ql}^{-1}+n\partial_{\mathcal{I}}^{2})\delta(\mathcal{I})\mathcal{I}^{2}\nonumber\\
&=&1-2J_{2}(K)-2J_{1}^{2}(K)+2J_{2}^{2}(K)+\ldots,
\end{eqnarray}
for $n\gg1$, where $\mathcal{I}\equiv I-I_{0}$. The terms of the series expansion (\ref{dnor}) coincides with the Rechester {\it et al.} results \cite{Rechester}, noting that $-J_{2}(2K)J_{4}(2K)\approx J_{2}^{2}(K)$ for $K\gg1$. Further corrections can be obtained by increasing the $k$-piece.

\begin{figure}[ht]
\resizebox{1.0\linewidth}{!}{\includegraphics*{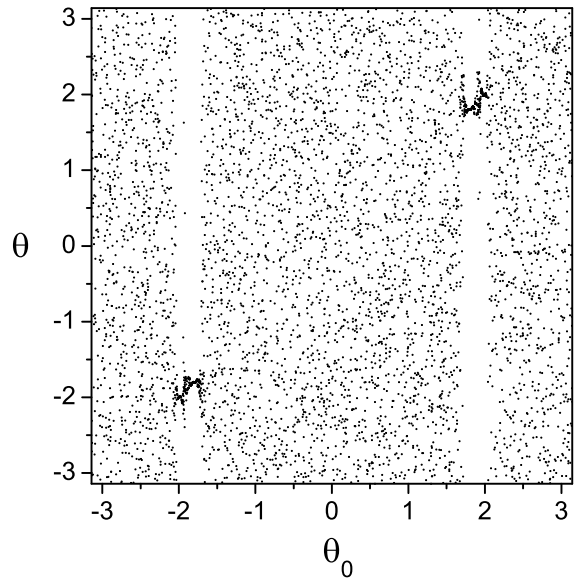}}
\caption{Measurements of $\theta$ (n=100) versus $\theta_{0}$ from numerical simulations performed for $N=5000$, $K=1.1\times2\pi$, $I_{0}=2\pi$, and $\theta_{0}$ uniformly distributed on $[-\pi,\pi)$. Accelerated orbits cluster along the diagonal $\theta=\theta_{0}$ on the gaps located inside the interval $\pi/2<|\theta_{0}|<\arccos(-4/K)\approx2.188$.}
\label{fig1}
\end{figure}

We turn now to the analysis of superdiffusion. In contrast to the dynamics of normal process, accelerator modes allow the PF operator $\hat{U}^{n}$ to load phase terms of the type $\exp(iI\sum_{k}m_{k})$ without the need of previous condition $\sum_{k}m_{k}=0$. On the other hand, accelerator modes also contribute to the corrections of the quasilinear value, although much stronger than the quasilinear one. Therefore, the stripes of wavenumbers that represent the leading contributions for the superdiffusion are
\begin{eqnarray}
\label{strip2}
&\underbrace{(0,0)\ldots(0,0)}_{n-r-1}\underbrace{(0,m)}_{1}\underbrace{(m,m)\ldots(m,m)}_{r},&\\
\label{strip3}
&\overbrace{(m,m)\ldots(m,m)}^{n},&
\end{eqnarray}
corresponding to particular configurations of $PU^{n}Q$ and $QU^{n}Q$, respectively. Adding the contributions due to stripes (\ref{strip2}) and (\ref{strip3}), the density $\rho_{n}$ assumes the form
\begin{eqnarray}
\label{rhophase}
\rho_{n}(I,\theta|I_{0},\theta_{0})&=&\frac{1}{2\pi}\sum_{r=1}^{n}\sum_{m\neq0}e^{im[(\theta-\theta_{0})-rI]}\nonumber\\
&\,&\times\Phi_{r,m}(K)\hat{J}_{0}^{n}\delta(I-I_{0})
\end{eqnarray}
where, neglecting $\mathcal{O}(J^{2})$ terms and considering $n\gg1$, we have
\begin{equation}
\label{phidef}
\Phi_{r,m}(K)=\left\{ \begin{array}{ll}
      J_{-m}(mK),\qquad\qquad\qquad\,r=1,\\
      \mathcal{O}(J^{2}),\qquad\qquad\qquad\,2\leq r<n,\\
      \sum_{j=1}^{n}J_{0}(jmK),\qquad\qquad r=n.
                 \end{array}
                 \right.
\end{equation}
The leading terms in the Eq. (\ref{rhophase}) are phases restricted to the following conditions
\begin{eqnarray}
\label{ly}
I\equiv I_{n}=2p\pi,\qquad \theta_{0}=\theta\equiv\theta_{n},
\end{eqnarray}
for any integer $p$, including $I_{0}=2p\pi$ as an alternative solution due to delta function $\delta(I-I_{0})$. In other words, only initial conditions that start from the vicinity of $I_{0}=2p\pi$ (or final conditions that reach it) for any $\theta_{0}\in\mathcal{S}_{\theta}$ and finish as a $Q$-periodic orbit ($1\leq Q\leq n$) must contribute to the superdiffusive transport. This scenario includes not only accelerator modes, but also irregular trajectories that are dragged along them. Such combination of regular and irregular paths makes it difficult to know precisely the probability measure for angular variables $(\theta,\theta_{0})$. In this sense, the angular amplitude along the accelerated dynamics can be estimated as
\begin{eqnarray}
\label{rhocc}
\int_{acc}d\theta d{\theta_{0}}\,e^{im(\theta-\theta_{0})}\sim\frac{2}{m^{2}}\Delta\theta\Delta\theta_{0}=\frac{1}{m^{2}}\frac{16\pi}{K},
\end{eqnarray}
where the maximal width of stability $\Delta\theta_{0}$ is given by $\mathcal{S}_{\theta}$:
\begin{equation}
\label{dt}
\Delta\theta_{0}=\arccos(-4/K)-\pi/2\approx4/K,
\end{equation}
valid for $K>4$. The averaged phase (\ref{rhocc}) is justified considering that: $i$) its value is lower than $2\Delta\theta\Delta\theta_{0}$, where the factor $2$ is due to the existence of two gaps; and $ii$) the same integral must be proportional to $m^{-2}$. In Fig. \ref{fig1} we illustrate this scenario for $K=1.1\times2\pi$ ($Q=3$ accelerated mode \cite{Ishizaki}), showing two gaps with widths $\Delta\theta_{0}$ as foreseen by Eqs. (\ref{stab}) and (\ref{dt}). Note that orbits lying in the gaps are captured by condition (\ref{ly}) while the others develop Brownian motion.

The stability interval for accelerated modes in terms of the parameter $K$ is given by
\begin{equation}
\label{kwin}
\mathcal{S}_{K}: 2|l|\pi<K<2|l|\pi\sqrt{1+(2/l\pi)^{2}},
\end{equation}
obtained through Eqs. (\ref{accmod}) and (\ref{stab}) for $Q=1$. The stability windows $\mathcal{S}_{\theta}$ and $\mathcal{S}_{K}$ are maximal since they contain all other $Q>1$ corresponding stability intervals for each corresponding value of $l$ \cite{Chirikov,Lichtenberg}. We can now compare the weight coefficients $\Phi_{1,m}(K)$ and $\Phi_{n,m}(K)$ subject to $\mathcal{S}_{K}$. For $K\approx2|l|\pi$, the sum $\sum_{m\neq0}m^{-2}\Phi_{1,m}(K)$ can be estimated as follows:
\begin{equation}
\label{phi1}
\sum_{m\neq0}\frac{J_{-m}(mK)}{m^{2}}\approx\frac{1}{\sqrt{8\pi K}}\sum_{m=1}^{\infty}(-1)^{m+1}c_{m}\sim\frac{1}{\sqrt{K}}
\end{equation}
where $c_{m}\equiv(m-1/2)^{-5/2}-m^{-5/2}$, noting that this series is convergent by the alternating series criterium. On the other hand, $\Phi_{n,m}(K)$ can be assessed for $n\gg1$ using a Schloemilch series \cite{Magnus}:
\begin{equation}
\label{sch}
\sum_{j=1}^{\infty}J_{0}(jmK)=-\frac{1}{2}+\frac{1}{|mK|}+\sum_{j=1}^{l'}\frac{2}{\sqrt{(mK)^{2}-(2j\pi)^{2}}},
\end{equation}
valid on the interval
\begin{equation}
\label{schint}
2l'\pi<|m|K<2(l'+1)\pi,
\end{equation}
where $l'$ is a positive integer. Comparing the restriction (\ref{schint}) with the stability condition (\ref{kwin}) it is easy to see that, in the limit $n\rightarrow\infty$, the superdiffusion rate will diverge on $K=2|l|\pi$ by setting $l'=|ml|$. Furthermore, due to the strong decay of angular amplitudes (\ref{rhocc}), the leading wavenumbers are $m=\pm1$. Finally, the mean-squared displacement due to accelerated dynamics is calculated by $\left\langle \mathcal{I}^{2}(n)\right\rangle_{sup}=\int d\mathcal{I}\rho_{acc}(\mathcal{I},n)\mathcal{I}^{2}$ resulting for the diffusion
\begin{eqnarray}
\label{diffacc}
\frac{D_{sup}}{D_{ql}}&=&\frac{16}{K}\left[\sum_{m=1}^{\infty}\frac{1}{m^{2}}\left(J_{-m}(mK)+\sum_{j=1}^{n}J_{0}(jmK)\right)\right]\nonumber\\
&\approx&\frac{16}{K}\sum_{j=1}^{n}J_{0}(jK)\qquad \mbox{for}\,\,\,K\in\mathcal{S}_{K}
\end{eqnarray}
and zero otherwise.

\begin{figure}[ht]
\resizebox{1.0\linewidth}{!}{\includegraphics*{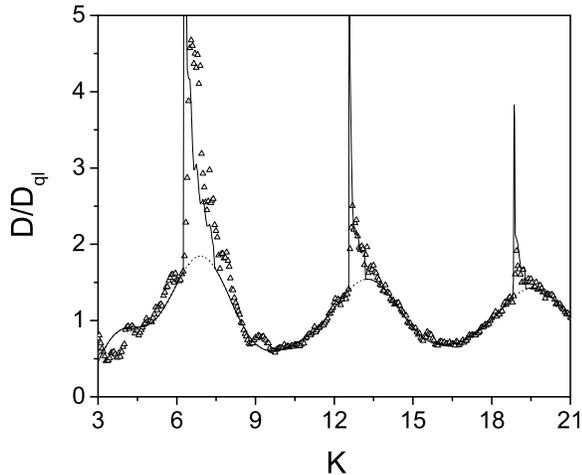}}
\caption{Theoretical diffusion rate $D/D_{ql}$ (solid line) compared with normal rate $D_{nor}/D_{ql}$ (dotted line) and measurements of $D/D_{ql}$ from numerical simulations ($\triangle$) for $n=100$ and $N=5000$. First and second theoretical diffusion peaks occur for values $14.63$ ($K=6.30$) and $6.07$ ($K=12.58$), respectively.}
\label{fig2}
\end{figure}

We have made extensive numerical studies of the superdiffusion coefficient (\ref{diffacc}), including the normal diffusion term (\ref{dnor}). A particular case for $n=100$ is showed in Fig. \ref{fig2}. It is clear that there is good agreement between theoretical and numerical results, especially when it is observed the accumulation of enhanced occurrences to the right of theoretical spikes in contrast to the their quasi absence on the left. In addition to the leading peaks located on the intervals $(\ref{kwin})$ with divergences at $K=2|l|\pi$ for $n\rightarrow\infty$, map (\ref{map}) has a more intrincated chain of secondary peaks whose phenomenology has been well studied in Ref. \cite{SM,Edelman,ZKBKWZ}. In order to describe them, we need to retain more terms in the evaluation of Eq. (\ref{rhophase}) including quasilinear stripes with a growing number of ``failures'' (sequence of different values for wavenumbers).

An important question concerns the determination of the superdiffusion exponent. As pointed out by Zaslavsky and Edelman \cite{Edelman}, typically it is hardly possible to get a theoretical value of $\beta$. If the power law $D_{sup}\propto n^{\beta-1}$ is asymptotically valid, then $\beta>1$ may exist only for values of $K$ for which $D_{sup}$ diverges. Using the asymptotic form of Bessel functions and noting that $\sum_{j=1}^{n}j^{-1/2}\sim2n^{1/2}$ for $n\gg1$, we can evaluate Eq. (\ref{diffacc}) for $K=2|l|\pi$ as
\begin{equation}
\label{beta}
\frac{D_{sup}}{D_{ql}}\sim\frac{n^{1/2}}{|l|^{3/2}}
\end{equation}
resulting $\beta=3/2$. In Ref. \cite{ZKBKWZ} there is a comparison between CTRW and FK formalisms applied to the study of superdiffusion of the map (\ref{map}). In both studies the numerical values of $\beta$ for $K=1.03084\times2\pi$ ($Q=5$ accelerated mode) are also very close to $3/2$, namely $\beta=1.42\pm0.15$. Note that Eq. (\ref{beta}) not only represents a class of universality for $\beta$ but also suggests the existence of other classes related to secondary diffusion peaks. In a similar way, further corrections for $D_{sup}$ may give secondary divergence terms for tighter $K$-intervals such that, together, they form a multifractal layer for the mean-squared displacement in the form 
$\left\langle I^{2}\right\rangle_{sup}\sim\sum_{j}C_{\beta_{j}}n^{\beta_{j}}$, where $\beta_{j}=\beta_{j}(K)$ as numerically observed in \cite{SM,Edelman,ZKBKWZ}.

The author thanks A. Saa, W.F. Wreszinski, and E. Abdalla for helpful discussions.


\begin{thebibliography}{99}
\bibitem{n1} For the unbounded momentum variable $I$ it is sufficient to assume a finite interval that includes some initial conditions $I_{0}=2p\pi$ due to periodicity of the map (\ref{map}).
\end{thebibliography}
\end{document}